\documentclass[sigchi]{acmart}

\newcommand{\mat}[1]{\mathbf{#1}}

\usepackage{amssymb,amsfonts}
\usepackage{subfig}
\usepackage{dblfloatfix}
\usepackage{mwe}

\usepackage{scalerel}

\copyrightyear{2020}
\acmYear{2020}
\setcopyright{acmcopyright}\acmConference[To Appear in ISWC '20]{Proceedings of the 2020 International Symposium on Wearable Computers}{September 12--16, 2020}{Mexico}
\acmBooktitle{Proceedings of the 2020 International Symposium on Wearable Computers (ISWC '20), September 12--16, 2020, Virtual Event, Mexico}
\acmPrice{15.00}
\acmDOI{10.1145/3410531.3414312}
\acmISBN{978-1-4503-8077-5/20/09}

\setcopyright{none}
\renewcommand\footnotetextcopyrightpermission[1]{}
\setcopyright{none}
\makeatletter
\renewcommand\@formatdoi[1]{\ignorespaces}
\makeatother

\begin{document}

\title{Towards Deep Clustering of Human Activities from Wearables}

\author{Alireza Abedin}
\affiliation{%
  \institution{The University of Adelaide}
}
\email{alireza.abedinvaramin@adelaide.edu.au}

\author{Farbod Motlagh}
\affiliation{%
  \institution{The University of Adelaide}
}
\email{farbod.motlagh@student.adelaide.edu.au}

\author{Qinfeng Shi}
\affiliation{%
  \institution{The University of Adelaide}
}
\email{javen.shi@adelaide.edu.au}

\author{Hamid Rezatofighi}
\affiliation{%
  \institution{The University of Adelaide}
}
\email{hamid.rezatofighi@adelaide.edu.au}

\author{Damith Ranasinghe}
\affiliation{%
  \institution{The University of Adelaide}
}
\email{damith.ranasinghe@adelaide.edu.au}

\renewcommand{\shortauthors}{Abedin et al.}

\begin{abstract}
  Our ability to exploit low-cost wearable sensing modalities for critical human behaviour and activity monitoring applications in health and wellness is reliant on supervised learning regimes; here, deep learning paradigms have proven extremely successful in learning activity representations from annotated data. However, the costly work of gathering and annotating sensory activity datasets is labor intensive, time consuming and not scalable to large volumes of data. While existing unsupervised remedies of deep clustering leverage network architectures and optimization objectives that are tailored for static image datasets, deep architectures to uncover cluster structures from raw sequence data captured by on-body sensors remains largely unexplored. In this paper, we develop an unsupervised end-to-end learning strategy for the fundamental problem of human activity recognition (HAR) from wearables. Through extensive experiments, including comparisons with existing methods, we show the effectiveness of our approach to jointly learn \textit{unsupervised representations} for sensory data and generate \textit{cluster assignments} with strong semantic correspondence to distinct human activities.
\end{abstract}


\maketitle

\section{Introduction}

Accurately and precisely understanding human activities is the basis for  applications ranging from assessing our cognitive decline, physical and mental health to performance in sporting activities~\cite{health6,mannini2017activity,chesser2019superlow,healthapp,pd,health1,health2,health4}. Increasing plethora of wearables are providing the opportunity to conveniently and at low-cost collect \textit{fine-grained} physiological information to understand human activities. However, the premise for realizing the multitude of applications is our ability to build accurate and, often personalized, models for recognizing human activities from wearables. 

\noindent\textbf{Problem.~} Human activity recognition problems have relied predominantly on supervised learning regimes where deep learning paradigms are extremely successful in learning activity representations from annotated data.  While the process of collection and annotation may be retrospective with vision based sensing modalities where visual inspections of, for example, video frames provides the basis for ground truth, the parallel task with wearables is nearly impossible. Moreover, such methods cannot be easily scaled to gather large datasets often necessary for deep neural networks (DNNs). In comparison to other domains, generating labelled data to benefit from supervised learning methods to build HAR applications in the absence of a reliable visualisation to establish ground truth is a unique HAR problem with wearable sensors. 

\noindent\textbf{Our Motivation.~} Although unsupervised methods provide avenues for learning from unlabelled data, investigations of unsupervised learning from wearable multi-channel time-series data remains dominantly limited to \textit{pre-training} \cite{alsheikh,deepautoset} or \textit{unsupervised representation learning} \cite{role,motion2vec}. 
Unsupervised alternatives without requiring any labels, such as \textit{deep clustering}, exist for 
 image data, 
however, these frameworks are tailored for still images and lack the \textit{inherent} capability to learn representations and clusters from raw sequential data captured by wearables. Therefore, our motivation is to investigate and develop a deep clustering architecture that:
\begin{itemize}
	\item Leverages the inherently sequential nature of sensory data.
	\item Learns \textit{clustering friendly} representations of activity features in the multi-sensor and multi-channel input signals that offer separability of activity classes in the feature space. 
	\item Promotes the formation of highly discriminative clusters with high semantic correspondence to human activities.
\end{itemize}

\noindent\textbf{Our Contributions.~} In this study, we propose \textit{Deep Sensory Clustering}---a deep clustering architecture that 
learns highly discriminative representations 
using 
self-supervision with \textit{reconstruction and future prediction tasks informed by feedback from a clustering objective to guide the network towards clustering-friendly representations}. The self-supervised tasks intend to incentivize the network to learn salient activity features that offer semantic separation in the feature space while simultaneously reducing the risk of collapsing clusters. Further, we augment the optimization objective with a clustering-oriented criterion to further refine the feature representations and gradually promote clustering-friendliness in the feature space. We validate our design concepts through extensive experiments; we summarize our key contributions below:

\begin{enumerate}
	\item We develop an \textit{unsupervised} deep learning network architecture for clustering human activities from raw sequences of wearable sensor data streams. Our approach, to the best of knowledge, provides the \textit{first} standalone, end-to-end, deep clustering method for \textit{raw} sequential data from wearables. 
	
	\item Through a systematic experimental regime conducted on three diverse HAR benchmark datasets (\texttt{UCI~HAR}, \texttt{Skoda}, \texttt{MHEALTH}), we demonstrate the effectiveness of our proposed approach. Further, we compare our method with closely related approaches, including traditional clustering methods. 
	
\end{enumerate}

\newpage

\section{Related Works}\label{related}


\paragraph{HAR with Wearable Sensors.}\label{HAR literature}  The superior performance of supervised deep neural networks in classification tasks has led to a shift towards the adoption of deep learning paradigms for recognizing human activities from raw wearable data \cite{sparsesense,nabil,misra,roy}. Researchers have explored CNNs \cite{convmobile,multichannelcnn,cnnphone,cnnnew,fcn}, RNNs \cite{bilstm,ensemble}, and a combination of convolutional and recurrent layers \cite{convlstm,attention,attend} to effectively model the temporal dependencies inherent in sequences captured with sensors. 
However, acquisition of labeled sensory data is labor-intensive and time-consuming. But, in the sequential sensor data domain, unsupervised learning has merely been investigated as a means for weight initialization \cite{alsheikh,role},  unsupervised feature learning prior to supervised fine tuning with labels \cite{motion2vec,deepautoset} or clustering of handcrafted features \cite{vade}, rather than a standalone end-to-end approach for exploiting cheaply accessible raw unlabelled data.

\begin{figure*}[t]
	\centering
	\includegraphics[width=0.6\textwidth, keepaspectratio]{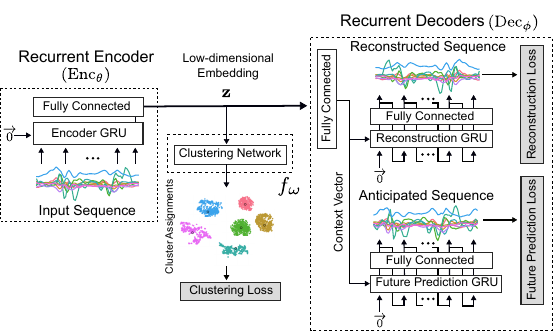}
	\caption{An Overview of \textit{Deep Sensory Clustering} pipeline.}
	\label{fig:pipeline}
\end{figure*}

\paragraph{Clustering with Deep Neural Networks.}
Recently, representation learning power of DNNs has been leveraged to achieve clustering-friendly representations and cluster assignments simultaneously for still image data; a shift towards \textit{Deep Clustering} paradigms \cite{cluster1}. In this regard, the feature space for representing images are initialized using deep autoencoders and iteratively refined to obtain cluster assignments \cite{dec,dbc}. Following similar ideas, Chen et al. \cite{dmc} propose a locality preserving criteria to learn structure preserving image representations, and Dizaji et al. \cite{depict} encourage balanced cluster assignments during training. In another study, a CNN is trained with agglomerative clustering objective in a recurrent process~\cite{jule}. Although these methods achieve impressive results for computer vision applications, existing deep clustering frameworks are tailored for still image datasets and  suffer from their inability to exploit the sequential nature of  wearable sensor data streams to learn representations and generate clusters of activities as substantiated in Section~\ref{clustering}.

\section{The Proposed Framework}\label{method}

We consider the problem of clustering a set of $n$ unlabelled segments of sensory readings $\{x_i\}^n_{i=1}$ into $k$ clusters, each representing a semantic human activity category. These segments are obtained by applying a sliding window of fixed temporal duration $\delta{t}$ over $d$ sensor channels of recorded datastreams. 
We propose our unsupervised two-staged \textit{Deep Sensory Clustering} framework illustrated in Fig. \ref{fig:pipeline} for the problem and detail our approach in what follows.

\subsection{Stage I: Pretraining a Multi-Task Autoencoder} 

In order to facilitate learning clustering-friendly representations, we initialize the feature space by pretraining a recurrent autoencoder to accomplish \textit{auxiliary tasks} in an unsupervised fashion. In accomplishing the delegated tasks, the network is \textit{forced to extract enriched representations} from the multi-channel sensor sequences. 


\paragraph{Recurrent Encoder ($\textrm{Enc}_{\theta}$).}
The encoder component of our network consumes a windowed excerpt of a raw multi-channel sensory sequence and learns a compact fixed length representation as a holistic summary of the input. In particular, we adopt a bi-directional GRU that reads through the partitioned sensory sequence $\mat{x}$ in both forward and backward directions and updates its internal hidden state in each time step according to the received input. The final hidden state obtained after scanning the entire input sequence is reduced in dimensionality through a fully connected layer. The resulting low-dimensional embedded feature, denoted by $\mat{z}\in\mathbb{R}^\mathrm{z}$, encodes contextual activity information by modeling the temporal dependencies present in the input sequence of sensory measurements $\mat{x}$.  We summarize the parameterized operations associated with encoding the input sequence $\mat{x}_i$ as $\mat{z}_i=\textrm{Enc}_{\theta}(\mat{x}_i)$. 


\paragraph{Conditional Recurrent Decoders ($\textrm{Dec}_{\phi}$).}
The decoder modules of the framework are structured symmetrically to the encoder component. First, a context vector is achieved by back projecting the embedded representation from the encoder into a higher-dimensional space such that it can be used to initialize the hidden states for the decoders. \textit{Two recurrent decoders then jointly exploit the generated context vector to accomplish different self-supervised tasks without requiring any manual supervision}. Inspired by \cite{Srivastava}, we share the encoder network between decoders with two different expertise; one decoder is specialized to reconstruct the temporally inversed input sequence, while the other one learns to \textit{anticipate the future sensory measurements} that should follow after, conditioned on the encoded input representation. Hence, the network has to not only learn a representation enriched with sufficient information to reproduce the input sequence, but also features that allow extrapolating future measurements. We summarize the parameterized decoding process as $(\mat{\hat{y}}_i^{\textrm{rec}},\mat{\hat{y}}_i^{\textrm{fut}})=\textrm{Dec}_{\theta}(\mat{z}_i)$, where $\mat{\hat{y}}_i^{\textrm{rec}}$ and $\mat{\hat{y}}_i^{\textrm{fut}}$ respectively denote the reconstructed and the anticipated sequences generated from the input $\mat{x}_i$ to satisfy the tasks.


\paragraph{Pre-training Objective.} We pre-train the entire recurrent autoencoder with a joint objective function,
\begin{equation}
\mathcal{L}_{\textrm{AE}}^{(i)}=\mathcal{L}_{\textrm{rec}}^{(i)} + \mathcal{L}_{\textrm{fut}}^{(i)}=\underbrace{\|\mat{y}_i^{\textrm{rec}}-\mat{\hat{y}}_i^{\textrm{rec}}\|^{\scaleto{2}{3pt}}}_{\text{reconstruction loss}} + \underbrace{\|\mat{y}_i^{\textrm{fut}}-\mat{\hat{y}}_i^{\textrm{fut}}\|^{\scaleto{2}{3pt}}}_{\text{future prediction loss}},
\label{eq:pretrain loss}
\end{equation}
where $\mathcal{L}_{\textrm{rec}}$ and $\mathcal{L}_{\textrm{fut}}$ denote the mean square error between each decoder's generated output sequence (\textit{i.e.}, $\mat{\hat{y}}^{\textrm{rec}}$ and $\mat{\hat{y}}^{\textrm{fut}}$) and the expected ground-truth target sequences (\textit{i.e.}, $\mat{y}^{\textrm{rec}}$ and $\mat{y}^{\textrm{fut}}$). Once the training is complete and the discrepancy between the generated outputs and their corresponding target sequences is minimized, the optimal network parameters, \textit{i.e.}, $(\theta^*, \phi^*)=\min_{\theta, \phi}\frac{1}{n}\sum_{i=1}^{n}\mathcal{L}^{(i)}_{\textrm{AE}}$, serve as an initialization point for the second stage. 

\subsection{Stage II: Representation Refinement with a Clustering Criterion} \label{refine}

Once the autoencoder becomes proficient in accomplishing the auxiliary tasks, we expect the feature space to find a semantic orientation. We further, extend our framework with a parameterized clustering network $f_\omega(.)$ capable of estimating cluster assignment distributions and iteratively optimize a clustering objective $\mathcal{L}_{\textrm{C}}$ to refine the feature space and guide the network towards yielding clustering-friendly representations. In this paper, we incorporate \textit{Cluster Assignment Hardening} \cite{dec} as a representative centroid-based approach for further refinement of the established feature space. During the refinement stage, both the clustering loss $\mathcal{L}_{\textrm{C}}$ and the autoencoding objectives $\mathcal{L}_{\textrm{AE}}$ are jointly incorporated to be optimized. Hence, the aggregated optimization criterion, for instance $i$, is formulated as
\begin{equation}
\mathcal{L}^{(i)}=\gamma\mathcal{L}_{\textrm{C}}^{(i)} + \mathcal{L}_{\textrm{AE}}^{(i)},
\label{eq:total_loss}
\end{equation} 
where the coefficient $\gamma\in[0,1]$ controls contribution of the clustering objective. Notably, we chose not to discard the decoding tasks during the refinement step to preserve the local data structure and allow a smoother manipulation of the feature space without distorting the previously established one. Once the network parameters are optimized with respect to the global criterion, $(\theta^*, \phi^*, \omega^*)=\min_{\theta, \phi, \omega}\frac{1}{n}\sum_{i=1}^{n}\mathcal{L}^{(i)}$, the clustering network of our framework directly delivers cluster assignments \textit{without} requiring a separate clustering algorithm to be run on the representations in a decoupled process. We describe the clustering criterion utilized next.


\paragraph{Cluster Assignment Hardening (CAH)} This clustering objective leverages the similarities between the data representations and the cluster centroids as a kernel to compute soft cluster assignments. Placing emphasis on the high confidence assignments, it then purifies the clusters and forces the assignments to have stricter probabilities. To incorporate this method, our clustering network $f_\omega$ comprises a single layer which maintains the cluster centroids $(\omega_j\in\mathbb{R}^\mathrm{z})_{j=1}^{k}$ as tunable network parameters and generates assignment distributions $Q_i=f_{\omega}(\mat{z}_i)$ for each instance $i$. This layer follows the Student's t-distribution to measure the similarity of embedded sequence representation $\mat{z}_i\in\mathbb{R}^\mathrm{z}$ to the $k$ cluster centroids and therefore, obtains the normalized similarities $Q_i=(q_{ij})^k_{j=1}$,
\begin{equation}
q_{ij}=\frac{(1+\|\mat{z}_i-\omega_j\|^{\scaleto{2}{3pt}})^{\scaleto{-1}{5pt}}}{\sum_{j'=1}^{k}(1+\|\mat{z}_i-\omega_j'\|^{\scaleto{2}{3pt}})^{{\scaleto{-1}{5pt}}}}.
\label{eq:q}
\end{equation}
Through squaring this distribution and then normalizing it, an auxiliary target distribution $P_i=(p_{ij})_{j=1}^k$ that leverages high confidence assignments is then defined to point the learning process towards stricter cluster assignments.
\begin{equation}
p_{ij}=\frac{q_{ij}^2/\sum_{i=1}^{n}q_{ij}}{\sum_{j'=1}^{k}(q_{ij'}^{2}/\sum_{i=1}^{n}q_{ij'})}.
\label{eq:p}
\end{equation}
Subsequently, the soft assignment distribution $Q_i$ is iteratively purified through minimizing the Kullback-Leilbler (KL) divergence between the soft labels and the auxiliary target distribution via training the network parameters, $\mathcal{L}^{(i)}_{\textrm{C}}=\textrm{KL}(P_i||Q_i)=\sum_{j=1}^{k}p_{ij}\log\frac{p_{ij}}{q_{ij}}.$
This centroid-based approach requires the cluster centers to be initialized \textit{once} at the beginning of the refinement stage. The initial centers are obtained by applying classical clustering algorithms on the acquired representations from the optimal pretrained parameters; \textit{i.e.}, $\{\mat{z}_i=\textrm{Enc}_{\theta^*}(\mat{x}_i)\}_{i=1}^{n}$.

\section{Experiments}\label{experiments}

\begin{table}[t]
	\centering
	\footnotesize
	\caption{A summary of the datasets explored in this work.}
	\resizebox{\columnwidth}{!}{%

		\begin{tabular}{lccc}
			\textbf{Dataset}            & \textbf{UCI HAR} & \textbf{Skoda} & \textbf{MHEALTH} \\ 
			\toprule
			Sensor Sampling Rate          & 50Hz    & 33Hz  & 50Hz    \\ 
			Sliding Window Duration ($\delta{t}$)      & 2.56s   & 1s    & 2.56s   \\ 
			Number of Sensor Channels ($d$)       & 9       & 60    & 23      \\ 
			Number of Activity Categories ($k$)     & 6       & 10    & 12      \\
			Number of Training Segments & 7352    & 5448  & 4088    \\ 
			Number of Testing Segments  & 2947    & 718   & 1022    \\  
		\end{tabular}}%

		\label{tab:datasets}
	\end{table}
	
\begin{table*}[t]
	\centering

	\footnotesize
	\caption{A quantitative comparison of clustering performance on three HAR benchmark datasets in accuracy (ACC) and NMI.}\label{tab:clustering}
	\resizebox{\textwidth}{!}{%
		\begin{tabular}{c||cccc||cccc||cccc}
			& \multicolumn{4}{c||}{UCI HAR Dataset}                               & \multicolumn{4}{c||}{Skoda Dataset}                                 & \multicolumn{4}{c}{MHEALTH Dataset}                               \\ 
			& \multicolumn{2}{c}{Train Split} & \multicolumn{2}{c||}{Test Split} & \multicolumn{2}{c}{Train Split} & \multicolumn{2}{c||}{Test Split} & \multicolumn{2}{c}{Train Split} & \multicolumn{2}{c}{Test Split} \\ 
			& NMI             & ACC            & NMI            & ACC            & NMI             & ACC            & NMI            & ACC            & NMI             & ACC            & NMI            & ACC            \\ \toprule
			\multicolumn{13}{l}{\textbf{Traditional Clustering on Input Data Space}}                                                                                                                                                                                                                           \\ \toprule
			\multicolumn{1}{l||}{$k$-means}                    & 44.28\%           & 48.25\%          & 42.28\%          & 42.14\%          & 43.41\%            & 41.01\%          & 46.01\%             & 40.67\%          & 49.71\%           & 39.55\%          & 48.37\%          & 42.37\%          \\ 
			\multicolumn{1}{l||}{AC-Average}                       & 1.38\%            & 19.16\%          & 1.93\%           & 18.29\%          & 4.61\%            & 14.34\%          & 30.98\%          & 26.04\%          & 4.44\%            & 9.21\%            & 7.55\%           & 9.31\%            \\ 
			\multicolumn{1}{l||}{AC-Complete}                       & 3.97\%            & 19.56\%          & 20.04\%          & 31.69\%          & 30.85\%           & 27.48\%          & 39.01\%             & 37.47\%          & 16.42\%           & 11.82\%          & 17.84\%          & 11.15\%          \\ 
			\multicolumn{1}{l||}{AC-Ward}                      & 41.07\%           & 42.26\%          & 48.21\%           & 43.26\%          & 46.55\%           & 44.68\%          & 46.92\%          & 41.78\%          & 54.06\%           & 45.16\%          & 56.99\%          & 45.99\%          \\ \toprule
			\multicolumn{13}{l}{\textbf{Traditional Clustering on Autoencoding Space}}                                                                                                                                                                                                                         \\ \toprule
			\multicolumn{1}{l||}{$k$-means}                    & 51.93\%           & 60.19\%          & 45.49\%          & 55.62\%          & 53.75\%           & 47.56\%          & 50.64\%          & 42.62\%          & 54.86\%           & 43.96\%          & 55.75\%          & 48.24\%          \\ 
			\multicolumn{1}{l||}{AC-Average}                       & 45.18\%           & 37.57\%          & 46.41\%           & 34.61\%          & 18.88\%           & 16.96\%          & 38.59\%          & 30.22\%          & 34.54\%           & 20.47\%          & 47.01\%          & 29.26\%          \\ 
			\multicolumn{1}{l||}{AC-Complete}                       & 40.66\%           & 40.03\%          & 40.81\%          & 43.67\%          & 32.55\%           & 32.47\%          & 41.57\%         & 35.93\%          & 42.23\%           & 35.05\%          & 44.42\%          & 36.51\%           \\ 
			\multicolumn{1}{l||}{AC-Ward}                      & 75.27\%           & 74.78\%          & 52.83\%          & 60.33\%          & 55.81\%           & 51.51\%          & 54.41\%          & 45.96\%          & 61.07\%           & 48.91\%           & 57.04\%          & 46.28\%          \\ \toprule
			\multicolumn{13}{l}{\textbf{End-to-End Deep Clustering}}                                                                                                                                                                                                                  \\ \toprule
			
			\multicolumn{1}{l||}{DEC \cite{dec}} & 
			52.85\%          & 50.45\%          & 53.00\%          & 49.85\%          & 45.32\%            & 40.46\%          & 47.06\%             & 40.25\%          
			& 51.86\%           & 43.64\% & 52.38\%          & 44.91\% \\ 
			
			
			\multicolumn{1}{l||}{IDEC \cite{idec}} & 54.86\%          & 51.14\%          & 50.47\%          & 50.15\%          & 49.54\%            & 47.41\%          & 47.47\%             & 45.96\%          & 50.89\%           & 42.49\% & 53.44\%          & 44.72\% \\ 

			\multicolumn{1}{l||}{(Ours) Deep Sensory Clustering ($k$-means Init.)} & 64.75\%          & 64.54\%          & 61.58\%          & 61.28\%          & 56.91\%            & 50.97\%          & 57.01\%             & 50.28\%          & \textbf{62.65\%}           & \textbf{57.19\%} & \textbf{63.06\%}          & \textbf{56.85\%} \\ 
			\multicolumn{1}{l||}{(Ours) Deep Sensory Clustering (Ward Init.)}      & \textbf{76.43}\%           & \textbf{78.79\% }         & \textbf{71.25\%}          & \textbf{75.41\%}           & \textbf{56.97\%   }        & \textbf{52.9\%}  & \textbf{59.06\%} & \textbf{53.48\%} & 59.42\%           & 51.57\%          & 60.91\%           & 53.33\%          \\ 
			
		\end{tabular}%
	}
	
\end{table*}

\begin{figure*}[t]
	
	\centering
	\includegraphics[width=\textwidth, keepaspectratio]{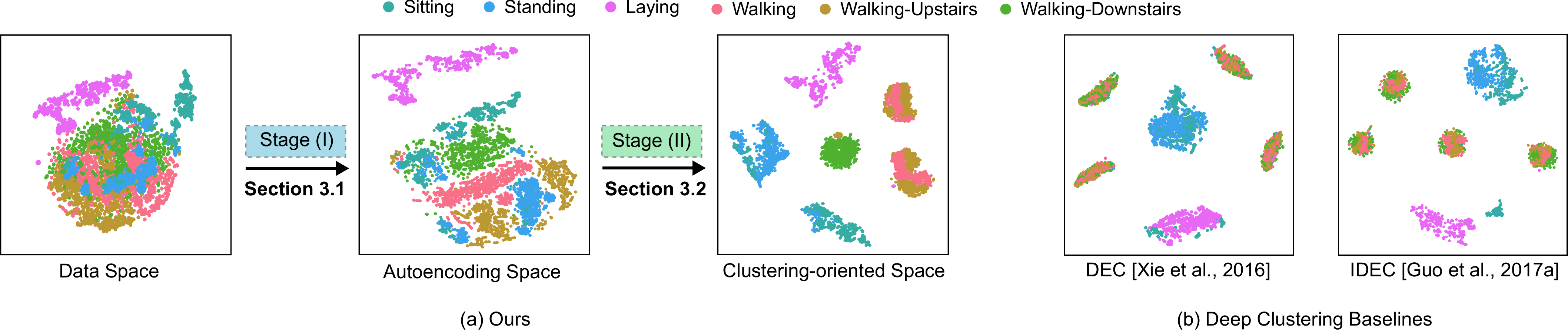}
	\caption{t-SNE visualizations of data representations for UCI HAR dataset achieved with (a) our proposed framework and, (b) deep clustering baselines. Sequence representations are color-coded with their corresponding ground-truth activity labels.}
	\label{fig:visual}
	
\end{figure*}

We ground our study by evaluating on three diverse HAR benchmark datasets: UCI HAR \cite{uci}; Skoda \cite{skoda}; and MHEALTH \cite{mhealth} employing standard train and holdout test splits (as summarized in Table~\ref{tab:datasets}). Datastreams are initially rescaled using per-channel normalization. After adopting the sliding window segmentation technique to partition the continuous data-streams, we consider the first $50\%$ of sensory measurements in each segment to constitute the input sequences to our framework. Accordingly, the temporally inversed version of the input is used as the target sequence for the reconstruction task while the remaining sensory measurements are considered as the target sequence for the future prediction task. 


\paragraph{Network Architecture.} We leverage a 2-layer bi-directional GRU with $256$ hidden units for the encoder. The decoders have an identical structure but utilize uni-directional connections. Considering the lower input dimension for UCI HAR compared with Skoda and MHEALTH datasets, we impose a bottleneck embedding dimension of $64$ for the former and $256$ for the latter in our autoencoders. The clustering network $f_\omega(.)$ for integrating CAH uses a single layer that generates soft cluster assignments according to Eq. (\ref{eq:q}). 


\paragraph{Optimization Settings.} In mini-batches of size $256$, the network parameters are updated using the ADAM optimizer with the initial learning rate set to $10^{-3}$ and decayed by a factor of $10$ after $70$ epochs. The network is pretrained for $100$ epochs, and fine-tuned with the clustering objective until the cluster assignment changes between two consecutive epochs is less than $0.1\%$. The weighting coefficient $\gamma$ is set to $0.1$. All above parameters are held constant across all datasets to refrain from unrealistic parameter tuning.
\subsection{Clustering}\label{clustering}

We base our evaluations for clustering on the two widely adopted metrics of \textit{unsupervised clustering accuracy} (\textit{ACC}) and Normalized Mutual Information (\textit{NMI}) \cite{cluster1}. Our approach is compared against popular centroid-based \textit{k}-means clustering \cite{kmeans++} as well as representative hierarchical algorithms including agglomerative clustering with average linkage (AC-Average) \cite{ac-average}, agglomerative clustering with complete linkage (AC-Complete) and Ward agglomerative clustering (AC-Ward). Further, we compare against end-to-end deep clustering methods proposed in~\cite{dec,idec} for still images and show their inability to cater for the sequential nature of time-series data.

\paragraph{Clustering Performance.} In Table \ref{tab:clustering}, we evaluate the clustering performance of the traditional baselines on both the:~$i)$~\textit{data space} using raw input representations; $ii)$ \textit{autoencoding space} using the embedded features $\{\mat{z}_i=\textrm{Enc}_{\theta^*}(\mat{x}_i)\}_{i=1}^{n}$ attained by optimizing $\mathcal{L}_{\textrm{AE}}$ in the pretraining stage; and $iii)$~compare with the \textit{end-to-end cluster assignments} generated by deep clustering baselines and our proposed \textit{Deep Sensory Clustering}. As required by the CAH objective, we report results over two different strategies to initialize the cluster centers \textit{only once} before commencing the refinement stage: $i)$~we run $k$-means clustering on the embedded features to obtain $k$ centroids; and $ii)$~we perform Ward clustering and use the mean representation of the obtained clusters as  initial centers.

Our results demonstrate that our end-to-end approach not only outperforms traditional clustering algorithms applied on both input data and auto-encoding spaces, but also offers a large performance margin over representative deep clustering baselines proposed for image data. Without any manual supervision, our proposed unsupervised approach can directly deliver cluster assignments with high correspondence to activities of interest in the explored datasets; we can observe accuracy (ACC) performance of $78.79\%$, $52.9\%$ and $57.19\%$, respectively on UCI HAR, Skoda and MHEALTH datasets. In addition, the \textit{consistent improvement} of unsupervised metrics across all three HAR datasets using our proposed framework demonstrates its generalizability to different HAR problems. 


\paragraph{Space Visualization.} In Fig. \ref{fig:visual}, we illustrate: \textit{i}) the evolution of the feature space towards the ultimate clustering-oriented embedding space achieved with our framework; and \textit{ii}) the deep clustering baselines by visualizing the data representation for the sequences in UCI HAR using t-SNE~\cite{tsne}. For our framework, we show the original dataset (\textit{data space}), the dataset embedded by the encoder after the pretraining stage (\textit{autoencoding space}) and the final representations after optimizing for the aggregated objective function $\mathcal{L}$ in Eq.~\ref{eq:total_loss} (\textit{clustering-oriented space}) with Ward initialization. We can observe that \textit{our framework discovers well-defined and clearly separated clusters of activity segments with strong correspondence to the ground-truth labels without manual supervision}. In contrast, the feature spaces achieved by the baseline deep clustering methods fail to correctly discover activity clusters; \textit{e.g.} static activities of \textit{sitting} and \textit{standing} are recognized as a single cluster, and different \textit{walking} variations are completely intermingled. These visualizations highlight: \textit{i}) necessity to leverage recurrent structures within the network;  and \textit{ii}) effectiveness of incorporating self-supervised tasks when dealing with time-series data from wearables.

\section{Conclusions}\label{conclusions}
This study tackles the hitherto unexplored problem of \textit{end-to-end clustering} of human actions from raw unlabelled multi-channel time-series data captured by wearables using a deep learning paradigm. To the best of knowledge, ours is the \textit{first} to investigate and develop a novel deep clustering architecture for HAR problems with raw sensor data. Our systematic experiments demonstrate: \textit{i}) the effectiveness; and \textit{ii}) generalizability of our proposed approach for clustering of human activities across three diverse HAR benchmark datasets. We believe our study creates new opportunities for recognition of human activities from unlabelled raw data that can be conveniently and cheaply collected from wearables.

\bibliographystyle{ACM-Reference-Format}
\bibliography{main}

\end{document}